\documentclass{aa}

\usepackage{graphicx}
\usepackage{txfonts}
\usepackage{gensymb}
\usepackage{wasysym}
\usepackage{caption}
\usepackage{subcaption}
\usepackage{threeparttable}
\usepackage{epstopdf}
\usepackage[normalem]{ulem}

\AppendGraphicsExtensions{.gif}

\usepackage{natbib}
\bibpunct{(}{)}{;}{a}{}{,}

\usepackage[dvipsnames]{xcolor}
\usepackage{hyperref}
\hypersetup{
    colorlinks=true,
    linkcolor=MidnightBlue,
    filecolor=blue,      
    urlcolor=cyan,
    citecolor=TealBlue
}

\begin{document}

   \title{From binary to singular: the AGN PSO J334.2028+1.4075 under the high-resolution scope}

   \author{P. Benke\inst{1,} \inst{2}
          \and
          K.~\'E.~Gab\'anyi\inst{3,} \inst{4,} \inst{5,} \inst{6}
          \and
          S.~Frey\inst{5,} \inst{6,} \inst{7}
          \and
          T.~An\inst{8,14}
          \and
          L.~I.~Gurvits\inst{9,} \inst{10}
          \and
          E.~Kun\inst{11,} \inst{12,}\inst{13,} \inst{5,}\inst{6}
          \and
          P.~Mohan\inst{8}
          \and
          Z.~Paragi\inst{9}
          \and
          E.~Ros\inst{1}
          }

   \institute{Max-Planck-Institut f\"ur Radioastronomie, Auf dem H\"ugel 69, D-53121 Bonn, Germany
         \and
             Institut f\"ur Theoretische Physik und Astrophysik, Universit\"at W\"urzburg, Emil-Fischer-Str. 31, D-97074 W\"urzburg, Germany
        \and 
            Department of Astronomy, Institute of Physics and Astronomy, ELTE E\"otv\"os Lor\'and University, P\'azm\'any P\'eter s\'et\'any 1/A, H-1117 Budapest, Hungary
        \and
            ELKH-ELTE Extragalactic Astrophysics Research Group, E\"otv\"os Lor\'and University, P\'azm\'any P\'eter s\'et\'any 1/A, H-1117 Budapest, Hungary
        \and 
             Konkoly Observatory, ELKH Research Centre for Astronomy and Earth Sciences, Konkoly Thege Mikl\'os \'ut 15-17, H-1121 Budapest, Hungary
        \and
             CSFK, MTA Centre of Excellence, Konkoly Thege Mikl\'os \'ut 15-17, H-1121 Budapest, Hungary
        \and
             Institute of Physics and Astronomy, ELTE E\"otv\"os Lor\'and University, P\'azm\'any P\'eter s\'et\'any 1/A, H-1117 Budapest, Hungary
        \and
            Shanghai Astronomical Observatory, Chinese Academy of Sciences, Shanghai, 200030, People's Republic of China
        \and
            Joint Institute for VLBI ERIC, Oude Hoogeveensedijk 4, 7991 PD Dwingeloo, The Netherlands
        \and
            Faculty of Aerospace Engineering, Delft University of Technology, Kluyverweg 1, 2629 HS Delft, The Netherlands
        \and
            Theoretical Physics IV, Faculty for Physics and Astronomy, Ruhr University Bochum, D-44780 Bochum, Germany
        \and
            Astronomical Institute, Faculty for Physics and Astronomy, Ruhr University Bochum, D-44780 Bochum, Germany
        \and
            Ruhr Astroparticle and Plasma Physics Center, Ruhr-Universität Bochum, D-44780 Bochum, Germany
        \and
            Key Laboratory of Radio Astronomy and Technology, Chinese Academy of Sciences, A20 Datun Road, Chaoyang District, Beijing, 100101, P. R. China
             }

   \date{Received ; accepted }


  \abstract{PSO\,J334.2028+1.4075 (PSO\,J334) is a luminous quasar located at redshift $z = 2.06$. The source gained attention when periodic flux density variations were discovered in its optical light curve. These variations were initially interpreted as the variability due to the orbital motion of a supermassive black hole binary (SMBHB) residing in a single circumbinary accretion disk. The orbital separation was determined to be $0.006$~pc with an in-spiral time of $7$\,yr in the rest frame of PSO\,J334. These findings suggested the quasar could be in the gravitational wave emitting phase of its merger and so extended multi-wavelength observations were commenced. However, subsequent observations provided evidence against the binary hypothesis as no optical periodicity was found on extended time baselines. On the other hand, detailed radio analysis with the Karl G.\ Jansky Very Large Array (VLA) and the Very Long Baseline Array (VLBA) revealed a lobe-dominated quasar at kpc scales, and possibly a precessing jet, which could retain PSO\,J334 as a binary SMBH candidate.}
  {We aim to study both the large- and small-scale radio structures in PSO\,J334 to provide additional evidence for or against the binary scenario.}
  {We observed the source at $1.7$\,GHz with the European Very Long Baseline Interferometry Network (EVN), and at $1.5$ and $6.2$\,GHz with the VLA, at frequencies that complement the previous radio interferometric study.}
  {Our images reveal a single component at parsec scales slightly resolved in the southeast–northwest direction and a lobe-dominated quasar at kiloparsec scales with a complex structure. The source morphology and polarization in our VLA maps suggest that the jet is interacting with dense clumps of the ambient medium. While we also observe a misalignment between the inner jet and the outer lobes, we suggest that this is due to the restarted nature of the radio jet activity and the possible presence of a warped accretion disk rather than due to the perturbing effects of a companion SMBH.}
  {Our analysis suggests that PSO\,J334 is most likely a jetted AGN with a single SMBH, and there is no clear evidence of a binary SMBH system in its central engine.}


   \maketitle
%

\begin{figure*}[h!]
    \centering
    \includegraphics[width=\linewidth]{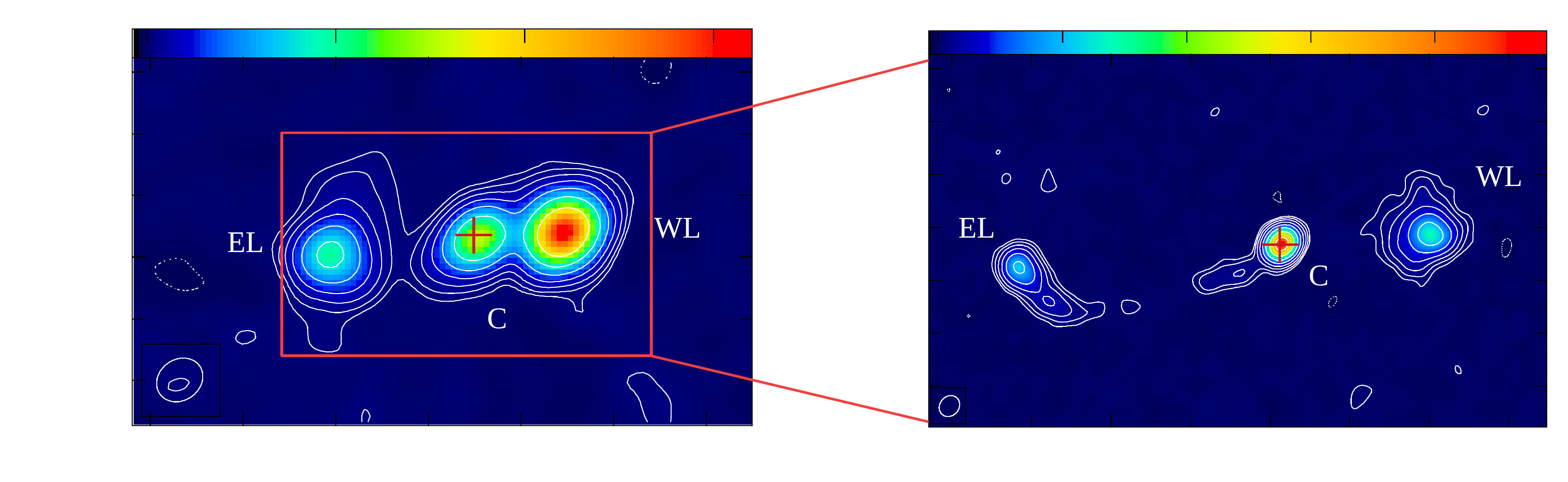}
    \caption{VLA images of PSO\,J334 at $1.5$ and $6.2$\,GHz. Each color bar represents total intensity in mJy\,beam$^{-1}$. Restoring beam sizes (FWHM) are $1218\,\mathrm{mas} \times 1665\,\mathrm{mas}$ at the major axis position angle $\mathrm{PA}=-30.5\degr$ at $1.5$\,GHz, and $340.6\,\mathrm{mas} \times422.4\,\mathrm{mas}$ at $\mathrm{PA}=-25.4\degree$ at $6.2$\,GHz, respectively. The lowest contours are at $0.12$ and $0.03$~mJy\,beam$^{-1}$, respectively, and further contour levels increase by a factor of two. The eastern and western lobes are marked as EL and WL, and the core is denoted as C. The red cross marks the position of the VLBI core at $1.7$\,GHz.}
    \label{fig:vla}
\end{figure*}

\begin{table*}[h]
         \centering
         \caption[]{Map properties of the clean images shown in Figs.~\ref{fig:vla} and \ref{fig:evn}, as well as the re-analyzed VLBA images from \citet{mooley18}.}
         \label{tab:img}
         \begin{tabular}{p{0.12\linewidth}ccccccc}
            \hline \hline
            \noalign{\smallskip}
            $\nu$ [GHz]\tablefootmark{a} & Array\tablefootmark{b} & $S_{\mathrm{tot}}$ [mJy]\tablefootmark{c} & $S_{\mathrm{peak}}$ [mJy\,beam$^{-1}$]\tablefootmark{d} & $\sigma$ [mJy\,beam$^{-1}$]\tablefootmark{e} & $b_{\mathrm{maj}}$ [mas]\tablefootmark{f}& $b_{\mathrm{min}}$ [mas]\tablefootmark{g} & PA [\degree]\tablefootmark{h}\\
            \noalign{\smallskip}
            \hline
            \noalign{\smallskip}
            1.5 & VLA & 35.41 & 14.7 & 0.04 & 1665 & 1218 & $-30.5$ \\
            6.2 & VLA & 11.48 & 4.52 & 0.01 & 422.4 & 340.6 & $-35.4$\\
            1.7 & EVN & 6.8 & 2.1 & 0.035 & 3.4 & 3.4 & $0$ \\
            4.38 & VLBA & 4.83 & 2.54 & 0.049 & 5.86 & 2.46 & $11.2$ \\
            7.40 & VLBA & 3.24 & 1.41 & 0.036 & 3.48 & 1.46 & $12.1$ \\
            8.67 & VLBA & 3.45 & 1.56 & 0.06 & 2.13 & 0.86 & $-0.27$ \\
            15.37 & VLBA & 1.19 & 0.85 & 0.036 & 1.29 & 0.48 & $-3.99$ \\
            \noalign{\smallskip}
            \hline
         \end{tabular}
         \tablefoot{
       \tablefoottext{a}{Observing frequency.}
       \tablefoottext{b}{Interferometer array performing the observation.}
       \tablefoottext{c}{Total flux density.}
       \tablefoottext{d}{Peak brightness.}
       \tablefoottext{e}{Rms noise.}
       \tablefoottext{f}{Beam major axis.}
       \tablefoottext{g}{Beam minor axis.}
       \tablefoottext{h}{Beam position angle.}
       }
\end{table*}

\section{Introduction}

The discovery of PSO\,J334.2028+1.4075 (FBQS\,J2216+0124; hereafter denoted as PSO\,J334) \citep{liu15} as a supermassive black hole binary (SMBHB) candidate attracted significant interest due to the rarity of confirmed SMBHBs. Supermassive black holes are expected to be at the center of most galaxies. Since galaxies evolve hierarchically, SMBHBs are believed to be abundant, especially at small separations \citep{an18}. However, confirming the existence of such objects has so far been mostly unsuccessful, with a few exceptions. The most notable examples are the dual AGN system in NGC\,6240 \citep{komossa03}, which resides in an ultraluminous infrared galaxy, and 0402+379, which was detected with the Very Long Baseline Array (VLBA) and has two radio cores separated by a projected distance of $7.3$~pc \citep{rodriguez06}. 

The SMBHB candidate quasar PSO\,J334 was discovered through a systematic search in the Pan-STARSS1 Medium Deep Survey \citep{liu15}. Based on an observed $542 \pm 15$ day period in the variation of the optical flux density and an estimated total black hole mass of $10^{9.97\pm 0.5} M_{\odot}$ (with a mass ratio between $0.05$ and $0.25$), an orbital separation of $0.006$~pc was inferred. According to this, the coalescence of the SMBHB would occur in approximately $7$~yr in the rest frame of the quasar. Unfortunately, current astronomical instruments are not capable of resolving the two components at such a small separation, so evidence for the existence of a second component could only be indirect, such as the detected periodic variability in the optical flux density. This variability could be caused by a secondary black hole passing through the primary black hole’s accretion disk, as proposed for OJ\,287 \citep{lehto96}. A similar explanation has been suggested in the case of the recently discovered SMBHB candidate SDSS\,J143016.05+230344.4 as well, which shows a periodic optical variability with a decay in both period and amplitude \citep{jiang22, an22}. However, the detected $2.6$ cycles of the putative periodicity are likely insufficient to claim sinusoidal variations \citep{vaughan16}, and other processes, such as quasi-periodic eruptions \citep{miniutti19} and quasi-periodic oscillations \citep{gierlinski08} may also explain the periodic flux density variability observed in PSO\,J334. Indeed, subsequent observations with extended time baselines of the optical monitoring failed to find any evidence of periodic variability in PSO\,J334 \citep{liu16}.

The radio structure of PSO\,J334 has been investigated in a multi-frequency study with the Karl G. Jansky Very Large Array (VLA) and the VLBA by \citet{mooley18}. In the VLBA images obtained at four frequencies from $4.4$ to $15.4$\,GHz, the quasar is resolved into two components, a compact core and a jet. Their separation is $3.6$ milliarcsec (mas), corresponding to a projected linear separation of $30$~pc. The VLA images at $2.8$ and $4.38$\,GHz reveal a lobe-dominated structure extending $66$~kpc from opposite sides of the core. In addition, the $39\degr$ difference between the position angles of the outer lobes and the inner jet is significant enough to suggest a perturbation of the jet by the second SMBH \citep{begelman80}, or alternatively a restarted double-double source. Thus, despite the results from the recent optical light curve, PSO\,J334 can still be considered as a SMBHB candidate \citep{mooley18}. Multi-wavelength observations aimed at determining the accretion mode of the quasar by \citet{foord17} have not found any feature that would convincingly distinguish PSO\,J334 from a single active galactic nucleus (AGN). However, there are still untested scenarios that would allow the object to retain its SMBHB status.

\begin{figure*}[h!]
    \centering
    \begin{subfigure}{0.49\linewidth}
        \includegraphics[width=\linewidth]{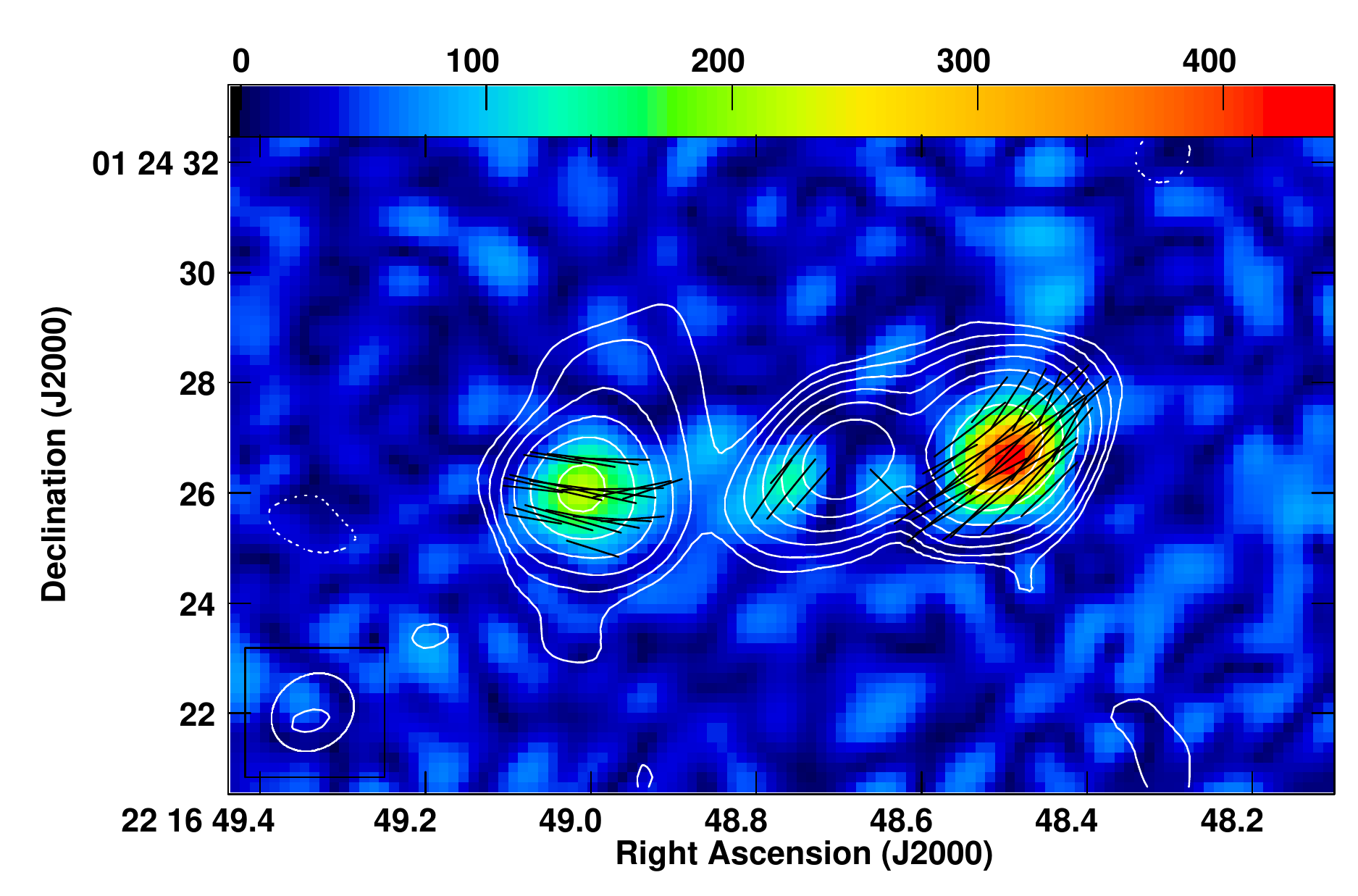}
    \end{subfigure}
    \begin{subfigure}{0.49\linewidth}
        \includegraphics[width=\linewidth]{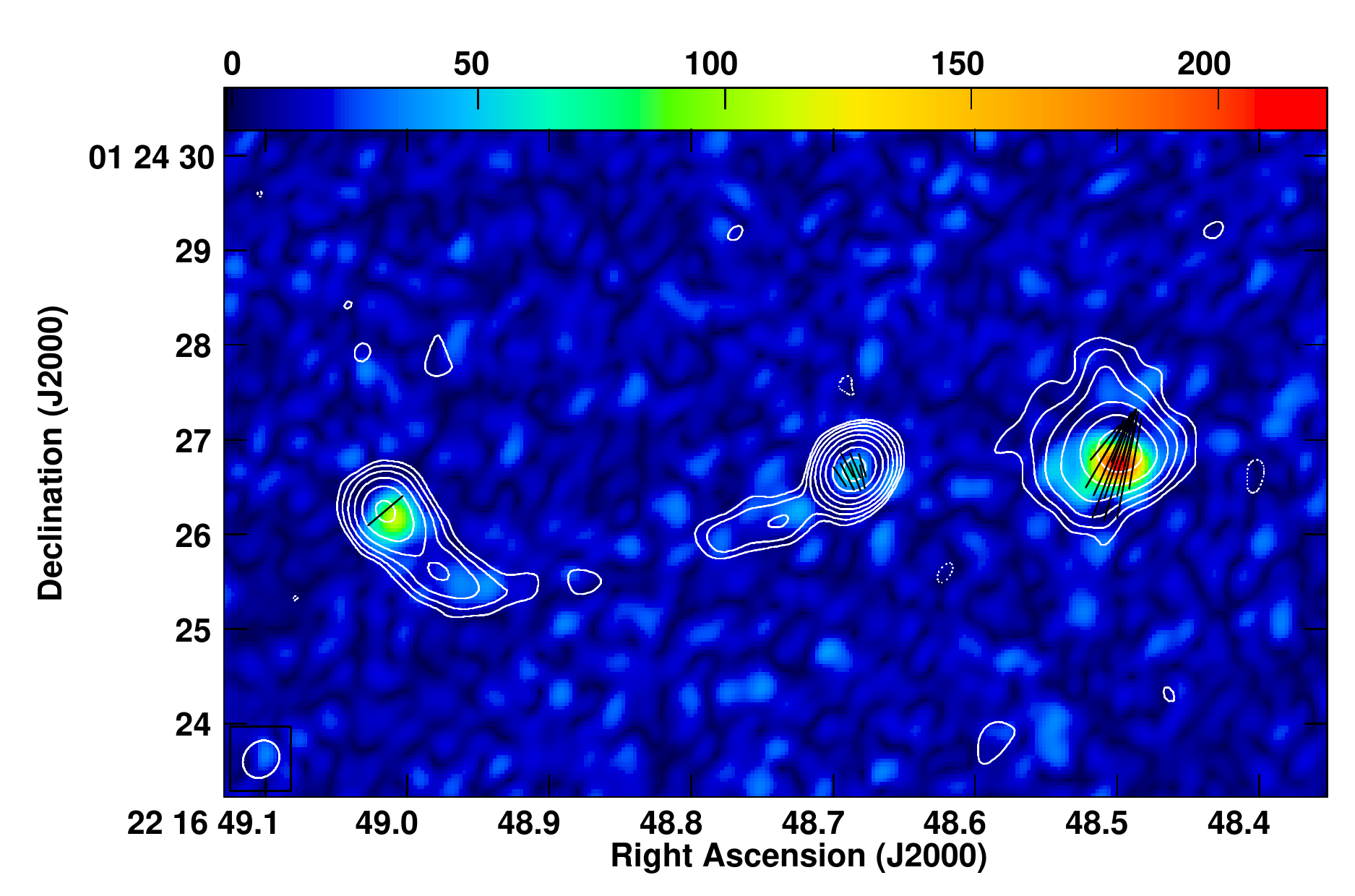}
    \end{subfigure}
    \caption{VLA $1.5$\,GHz (left panel) and $6.2$\,GHz (right panel) polarimetric images of PSO\,J334. The colors represent the polarized intensity in $\mu$Jy\,beam$^{-1}$, while the contours are the same as in the total intensity map in Fig.~\ref{fig:vla}. EVPAs are represented by the black ticks in the image.}
    \label{fig:pol}
\end{figure*}

We study the radio structure of PSO\,J334 using the technique of very long baseline interferometry (VLBI) with the European VLBI Network (EVN) at $1.7$~GHz and with the VLA at $1.5$ and $6.2$~GHz. Here we present our results and compare them with those obtained with the VLBA at higher frequencies by \cite{mooley18}. In Sect.~\ref{obs}, we describe the observations and the data reduction process. We then discuss our results in Sect.~\ref{res}. Finally, a summary is given in Sect.~\ref{sum}. 
In this paper we assume a $\Lambda$CDM cosmological model with $H_0=70$~km\,s$^{-1}$\,Mpc$^{-1}$, $\Omega_{\Lambda}$=0.73, and $\Omega_{\mathrm{M}}$=0.27. At the redshift of the object, $z=2.06$ \citep{becker01}, $1\arcsec$ of angular distance in the sky corresponds to $8.569$~kpc of projected linear distance \citep{wright06}.

\section{Observations and data reduction}
\label{obs}

\subsection{VLA data}
\label{vla}

Observations with the VLA (project code AG980, PI: K.\'E. Gab\'anyi) were carried out at $1.5$ and $6.2$\,GHz (L and S/C bands) on 2016 October 28 and 2016 October 26, respectively. The VLA was observing in its most extended A configuration with $27$ antennas. The on-source time was $0.5$\,h in both bands. The primary flux density calibrator in both experiments was 3C\,48, and the polarization D-term calibrator was \object{J2355+4950}. The secondary calibrators were \object{J2212+0152} ($1.5$\,GHz) and \object{J2218$-$0335} ($6.2$\,GHz). The $1.5$\,GHz data were recorded in $16$ spectral windows between $1.008$ and $1.968$~GHz with $64$ channels, each with a bandwidth of $64$~MHz. The $6.2$\,GHz data contained $48$ spectral windows, but the first $16$ were only used to set up the observations, so the target and calibrators were observed between $4.226$ and $8.096$~GHz frequencies in $32$ spectral windows, each with $64$ channels and a bandwidth of $128$~MHz.

Data reduction was performed in the Common Astronomy Software Applications \citep[\texttt{CASA},][]{mcmullin07,casa22} package version $7.15.0$. First, phase, delay, bandpass, and gain calibrations were derived for the primary calibrator. Then, to calibrate polarization, we determined cross-hand delays, solved antenna-based D-terms for the unpolarized calibrator, and finally calibrated the polarization angle for the primary calibrator. We then used the calibration tables generated in the previous steps to transfer the solutions to the secondary calibrator and then to our target source\footnote{For the calibration, we followed this VLA tutorial: \url{https://casaguides.nrao.edu/index.php/VLA_Continuum_Tutorial_3C391-CASA5.5.0}}. As 3C\,48 was undergoing an active phase during 2016\footnote{\label{note2}\url{https://science.nrao.edu/facilities/vla/docs/manuals/obsguide/modes/pol}}, we inspected the polarimetric calibration carefully, imaged all calibrators to determine the polarization angle and fractional polarization values, and compared them to those available in the literature. In the case of the D-term calibrators, we found no significant polarization signatures and fractional polarization values were close to $0$. In the case of 3C\,$48$, polarization angles and fractional polarization values in the two bands differ by less than $15$\% compared to the values in the literature\textsuperscript{\ref{note2}}.

Hybrid imaging was performed by iterating \texttt{tclean} and self-calibration, while imaging all four Stokes parameters together, and then deriving polarized intensity and polarization fraction images from the Stokes IQU images. The $1.5$ and $6.2$\,GHz VLA images are shown in Fig.~\ref{fig:vla}, and the polarized intensity and electric vector position angle (EVPA) images are shown in Fig.~\ref{fig:pol}.

\subsection{EVN data}
\label{evn}

Our EVN observation (project code RSG08, PI: S. Frey) at $1.7$~GHz was performed on 2015 October 18, with the participation of eleven radio telescopes: Jodrell Bank Lovell Telescope (United Kingdom), Westerbork (single dish; the Netherlands), Effelsberg (Germany), Medicina (Italy), Onsala (Sweden), Sheshan (China), Toru\'n (Poland), Hartebeesthoek (South Africa), Svetloe, Zelenchukskaya, and Badary (Russia). The data were recorded at a $1$~Gbit\,s$^{-1}$ rate in left and right circular polarizations, with 8 basebands (IFs) per polarization, each divided into thirty-two $500$-kHz wide spectral channels. The total bandwidth was $128$~MHz per polarization. The correlation was performed at the Joint Institute for VLBI ERIC (Dwingeloo, the Netherlands) with $4$~s integration time. The observation lasted for a total of $2$~h. We used phase-referencing to a nearby (separated by $0\fdg96$) compact calibrator source, J2217+0220, with duty cycles of $6.5$~min, including $3$~min  $20$~s scans spent on the target. The total accumulated observing time on PSO\,J334 was $1$~h. 

\begin{figure}[h!]
    \centering
    \includegraphics[width=0.95\linewidth]{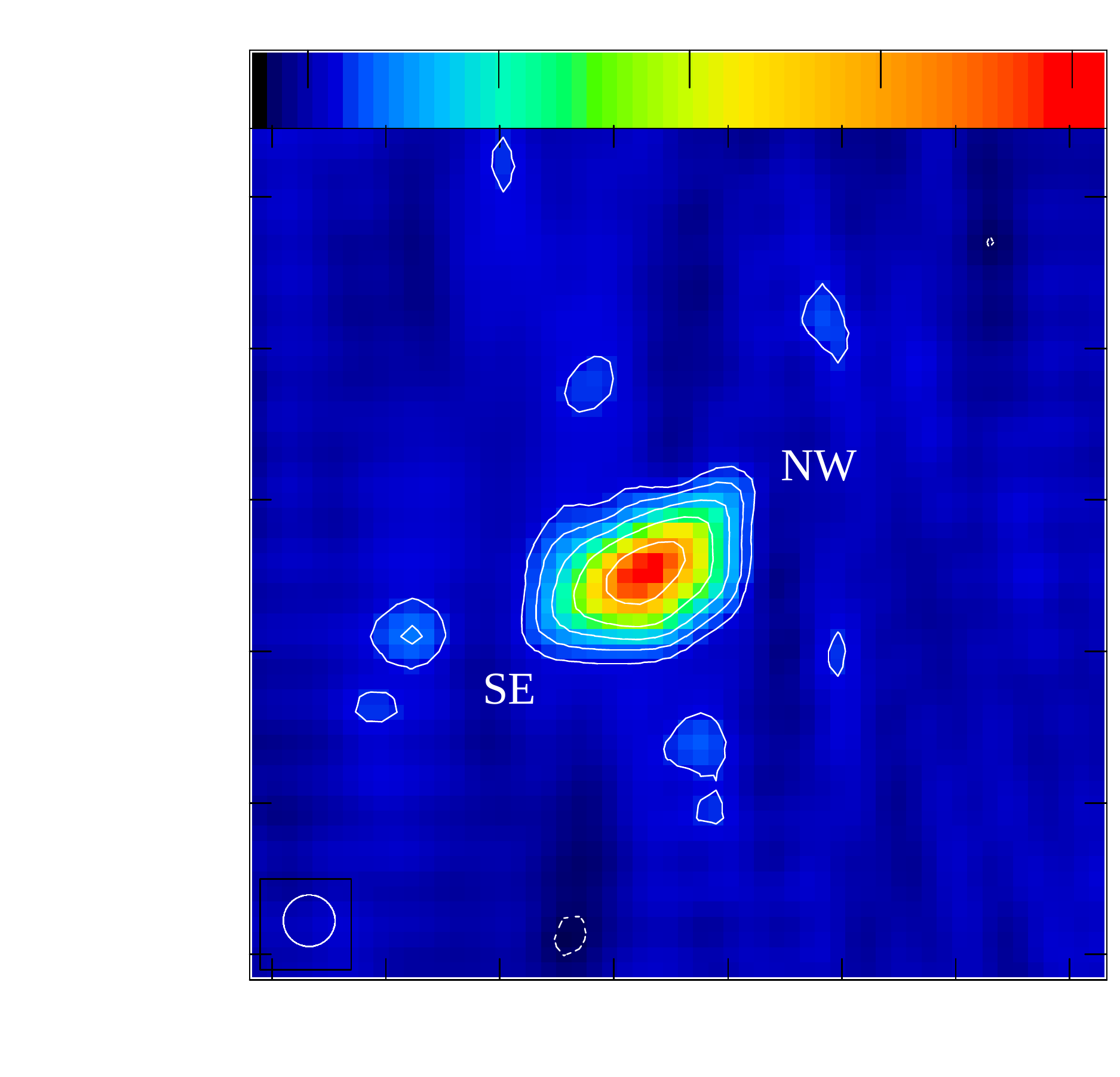}
    \caption{$1.7$-GHz EVN image of PSO\,J334. The image is restored with a $3.4$-mas circular Gaussian beam. The lowest contours are at $0.09$~mJy\,beam$^{-1}$ and increase by a factor of two. The source shows an elongated shape in the southeast--northwest direction, which is consistent with the structure reported in \citet{mooley18}.}
    \label{fig:evn}
\end{figure}

The data were calibrated with the U.S. National Radio Astronomy Observatory (NRAO) Astronomical Image Processing System \citep[\texttt{AIPS},][]{greisen03}, following the usual procedures. The amplitudes of the raw correlated visibility data were calibrated using the antenna gain curves and measured system temperatures (where available), as provided by the participating stations. Nominal system temperature values were used for Jodrell Bank, Sheshan, Svetloe, Zelenchukskaya, and Badary. The data were then corrected for the dispersive ionospheric delay by invoking the task \texttt{TECOR} that uses total electron content maps derived from global navigation satellite systems data. Phase changes due to the time variation of the source parallactic angle were also corrected for azimuth--elevation mounted radio telescopes in the network. Global fringe-fitting  was performed using the task \texttt{FRING} on the phase-reference calibrator J2217+0220 and the bright fringe-finder source J2148+0657 also observed for a $12$-min scan at the beginning of the experiment. These calibrated visibility data were exported to \texttt{Difmap} \citep{shepherd97} for imaging. Conventional hybrid mapping with several iterations of the \texttt{clean} algorithm \citep{högbom74} and phase-only self-calibration was performed. Then antenna-based gain correction factors were determined. These values were within $\pm5\%$ for the compact bright fringe-finder source, suggesting a reliable initial amplitude calibration. The \texttt{clean} component model obtained for the phase-reference source J2217+0220 in \texttt{Difmap} was fed back to \texttt{AIPS}, before repeating fringe-fitting, now taking the calibrator source structure into account for determining visibility phases. The fringe-fit solutions obtained for J2217+0220 were interpolated to the target source, PSO\,J334, within the atmospheric coherence time using the task \texttt{CLCAL}. The final calibrated visibility data file for PSO\,J334 was then transferred to \texttt{Difmap} for imaging and brightness distribution modeling.

The peak of the dirty image was offset by about $1\farcs1$ from the center because of an inaccurate a-priori position used for radio telescope pointing. Therefore we started the imaging by shifting the phase center to the actual brightness peak. Because PSO\,J334 is relatively weak with $\sim 7$~mJy total flux density, and it appeared slightly resolved at $1.7$~GHz with the EVN, self-calibration in general was not attempted during hybrid imaging in \texttt{Difmap}, except for phase self-calibration for the European stations (Effelsberg, Jodrell Bank, Onsala, Westerbork, Medicina, and Toru\'n). We used gradually decreasing solution intervals from $60$ to $1$~min. Amplitude self-calibration was not made at all. The naturally weighted \texttt{clean} image of PSO\,J334 is shown in Fig.~\ref{fig:evn}. A circular two-dimensional Gaussian brightness distribution model component fitted directly to the self-calibrated visibility data in \texttt{Difmap} can adequately describe the source, allowing us to quantitatively characterize its size and flux density (Table~\ref{tab:comp}). Neither an elliptical Gaussian, nor a two-component circular Gaussian model could significantly improve the goodness of fit.

\subsection{Archival VLBA data}
\label{vlba}

To be able to perform brightness distribution model fitting in the visibility domain, we downloaded from the NRAO data archive\footnote{\url{https://data.nrao.edu/}}, re-calibrated, and imaged the VLBA data of \citet{mooley18} measured at $4.38$, $7.40$, $8.67$, and $15.37$~GHz (project code BM438, PI: K.P. Mooley). For details on the observations, we refer to the original paper \citep{mooley18}. Calibration was carried out in \texttt{AIPS}. After loading the data with \texttt{FITLD}, we performed parallactic angle and digital sampling corrections, corrected for the Earth orientation parameters, and applied ionospheric corrections that are especially important for phase-referenced observations performed at low frequencies and low source declination. At first, fringe-fitting was performed on the phase-reference calibrator, J2217+0220, using the task \texttt{FRING}. After applying the calibration tables and writing the data out of \texttt{AIPS}, we performed hybrid imaging in \texttt{Difmap}, in a similar way as described in Sect.~\ref{evn} for the EVN data. Since the gain corrections for all the VLBA antennas were within $\pm$5\%, we did not perform any additional antenna-based amplitude correction in \texttt{AIPS}. The \texttt{clean} image of J2217+0220 was then loaded into \texttt{AIPS} and used during the second round of fringe-fitting. Delay and rate solutions were applied to both the calibrator and the target, PSO\,J334. The calibrated visibility data of the target source were written out. Imaging again was carried out in \texttt{Difmap} by only using \texttt{clean} iterations without self-calibration. The total flux densities, i.e. the sum of the individual clean components, and peak brightnesses agree with the values published in \citet{mooley18} within 10\% (see image properties in Tab.~\ref{tab:img}). Gaussian brightness distribution model components (Table~\ref{tab:comp}) were fitted to the visibility data using the \texttt{modelfit} command in \texttt{Difmap}.

\begin{figure*}[h!]
    \centering
    \begin{subfigure}{0.52\linewidth}
        \includegraphics[width=\linewidth]{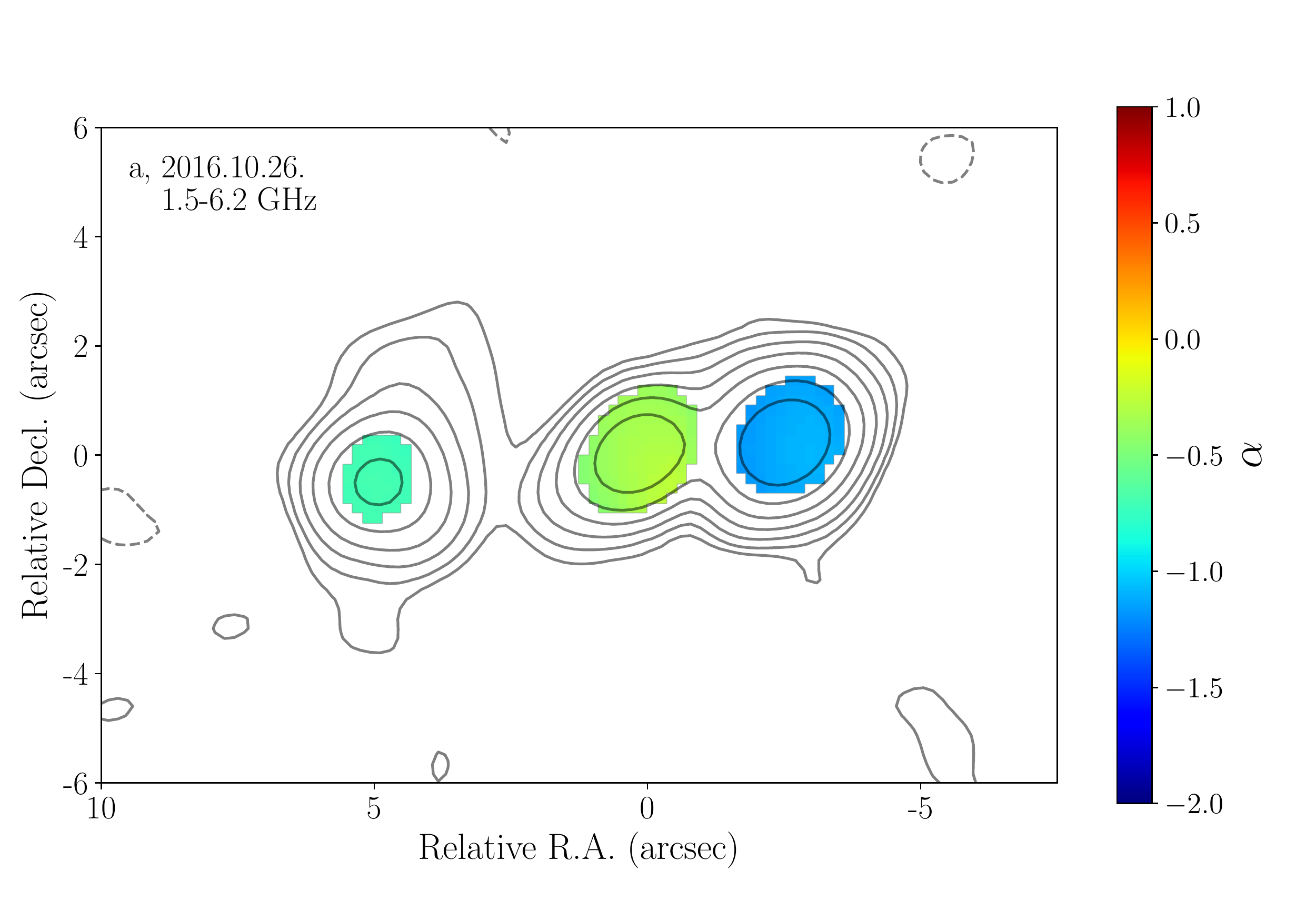}
    \end{subfigure}
    \begin{subfigure}{0.44\linewidth}
        \includegraphics[width=\linewidth]{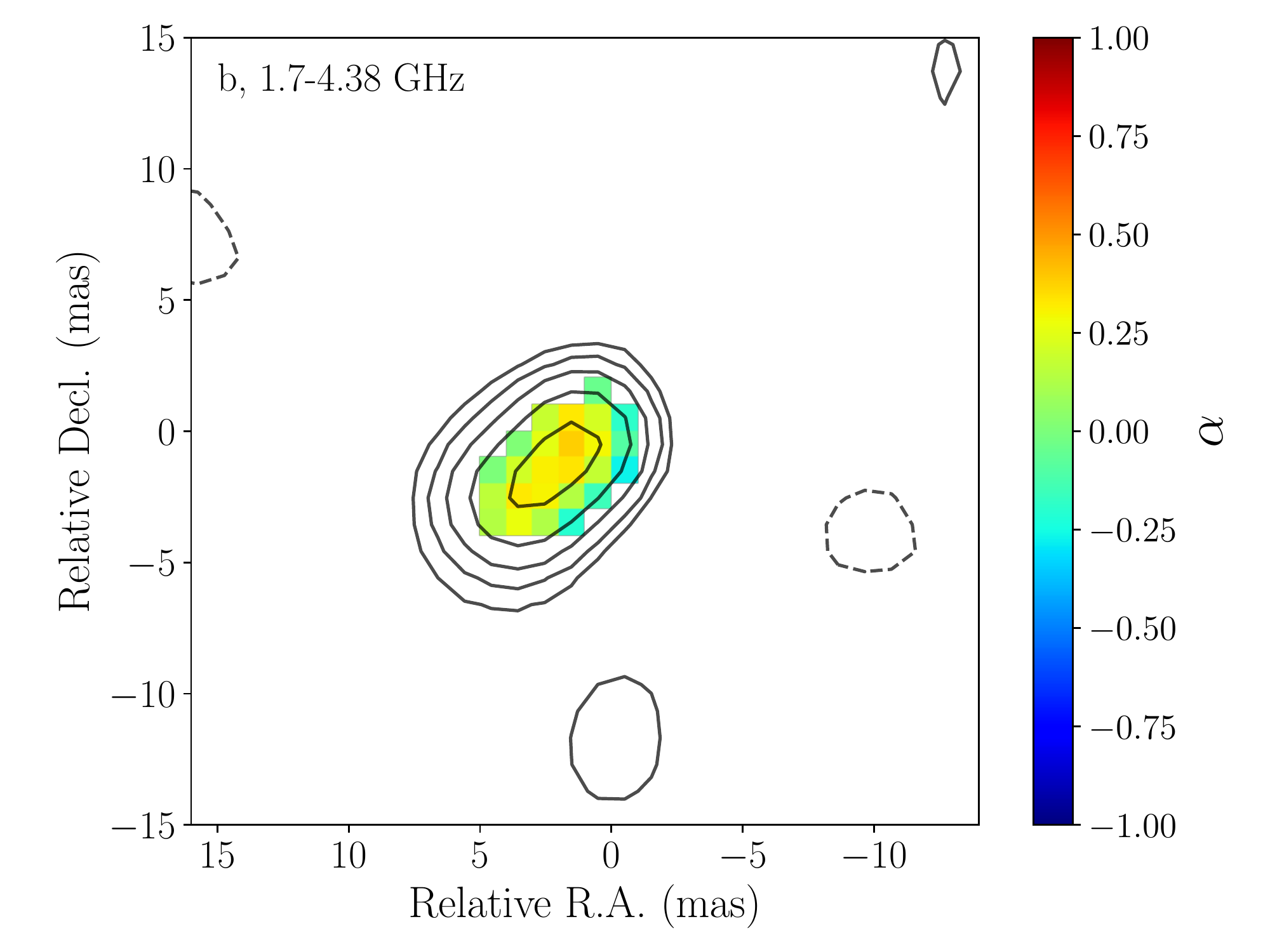}
    \end{subfigure}
    \caption{Two-point spectral index distribution maps of PSO J334 based on: \textit{(a)} quasi-simultaneous VLA images at $1.5$ and $6.2$\,GHz, and \textit{(b)} $1.7$-GHz EVN and $4.38$-GHz VLBA images. Lowest contours are at $0.12$  and $0.14$~mJy/beam, respectively, and contour levels increase by a factor of two. Colors represent spectral index values. We note that the VLBI images were taken half a year apart, so flux density variability cannot be excluded and therefore the spectral index map in panel \textbf{\textit{(b)}} should be treated with caution.}
    \label{fig:spec}
\end{figure*}

\section{Results and discussion}
\label{res}

\subsection{Source structure and polarization}
\label{subs:str-pol}

The $1.7$-GHz EVN image of PSO\,J334 (Fig.~\ref{fig:evn}) restored with a $3.4$-mas circular Gaussian beam (full width at half maximum, FWHM) shows a single component that is slightly resolved in roughly the southeast--northwest direction. The structure is consistent with the higher-frequency VLBA images \citep{mooley18}, where two components -- a southeastern synchrotron self-absorbed core and a northwestern jet -- were identified with $3.6$~mas separation along the position angle of $139\degr$. (Position angles are conventionally measured from north through east.) The $39\degr$ misalignment in position angles between the VLA and VLBA jets led \citet{mooley18} to restore the SMBHB status of PSO\,J334, however, alternative interpretations are still viable to explain the observations. Double-double radio sources, such as B0925+420 and B1450+333 \citep{schoenmaker00}, that retain signs of several active phases show similar morphology to PSO\,J334. In this case, the lobes seen on kpc scales with the VLA would be relics of past activity, and the compact, mas-scale core--jet morphology was formed more recently. The change in jet position angle can be interpreted as precession either due to interaction with a companion SMBH \citep{begelman80} or caused by a warped accretion disk changing the orientation of the jet \citep{pringle97}.

Our VLA observations at $1.5$ and $6.2$\,GHz show details of the complex structure of PSO\,J334 (Fig.~\ref{fig:vla}). The source, as shown in the $2.8$-GHz VLA B-configuration Caltech--NRAO Stripe $82$ Survey (CNSS) image of \citet{mooley18}, is a lobe-dominated quasar oriented at a large angle with respect to our line of sight. The projected linear size of the object is approximately $79$~kpc, which is typical of $z=2$ AGN \citep{blundell99}. The lengths of the arms are unequal, with the eastern arm being longer at both frequencies. However, contrary to expectations that the longer arm is pointing towards the observer and is brighter due to Doppler boosting \citep{longair79}, here the western, shorter arm appears brighter in both images. 
In addition, the eastern arm shows a sharp turn before ending in a hotspot, which suggests an interaction with the surrounding interstellar medium (ISM), similarly as, e.g., in the high-redshift radio galaxy 4C\,$41.17$ \citep{gurvits97}. However, we do not detect polarized emission at the turning point as, e.g., in PKS\,0637$-$752 \citep{lovell00}, where the highly polarized bent jet region also gives rise to bright X-ray emission \citep{schwartz00}. By inspecting the polarimetric images, we also see that the polarized intensity is higher in the western lobe and that EVPAs are close to perpendicular to the jet propagation on both sides, indicating the presence of a termination shock where the lobe material interacts with the ambient medium. The asymmetric structure together with the polarimetric results suggest that PSO\,J334 is embedded in a large-scale environment that is not intrinsically symmetric, and the jet interacts with clumps of the ISM that are disrupted upon contact with the jet material.

Inspecting the inner $2\arcsec-3\arcsec$ in our $6.2$-GHz VLA image (Fig.~\ref{fig:vla}), we see a remarkably straight jet that cannot be described with any precession model. Jet precession itself would be revealed by a helical jet shape that is physically external or intrinsic to the jet. In the first case, the main driver might be e.g. binary motion \citep[e.g.][]{kun2014}, Lense--Thirring precession \citep[e.g.][]{lensethirring,liska2018}, or disk precession induced by a secondary black hole \citep[e.g.][]{caproni2004}. In this case, the jet components move more or less on straight or ballistic paths, and the pitch of the jet is constant. The helical pattern simply reflects the periodic ejection direction of the newborn jet components. The second case appears, e.g., due to instabilities in the jet \citep[e.g.][]{perucho2006} and components actually move along helical path. In this case, the spatial wavelength of the jet along its symmetry axis is increasing with increasing core separation. The jet appearing in the $6.2$-GHz VLA image (Fig.~\ref{fig:vla}) has no resemblance to any of these scenarios. However, we cannot rule out the possibility that more sensitive images with higher angular resolution could recover jet structures indicative of precession or the presence of instabilities \citep[see, e.g., the case of 3C\,$279$,][]{fuentes22}. This structure seems hard to reconcile with the jet precession suggested by \citet{mooley18} based on lower-resolution VLA data, and rules out the last argument supporting the binary nature of the source.

\subsection{Spectral analysis}

\begin{figure}[h!]
    \centering
    \includegraphics[width=\linewidth]{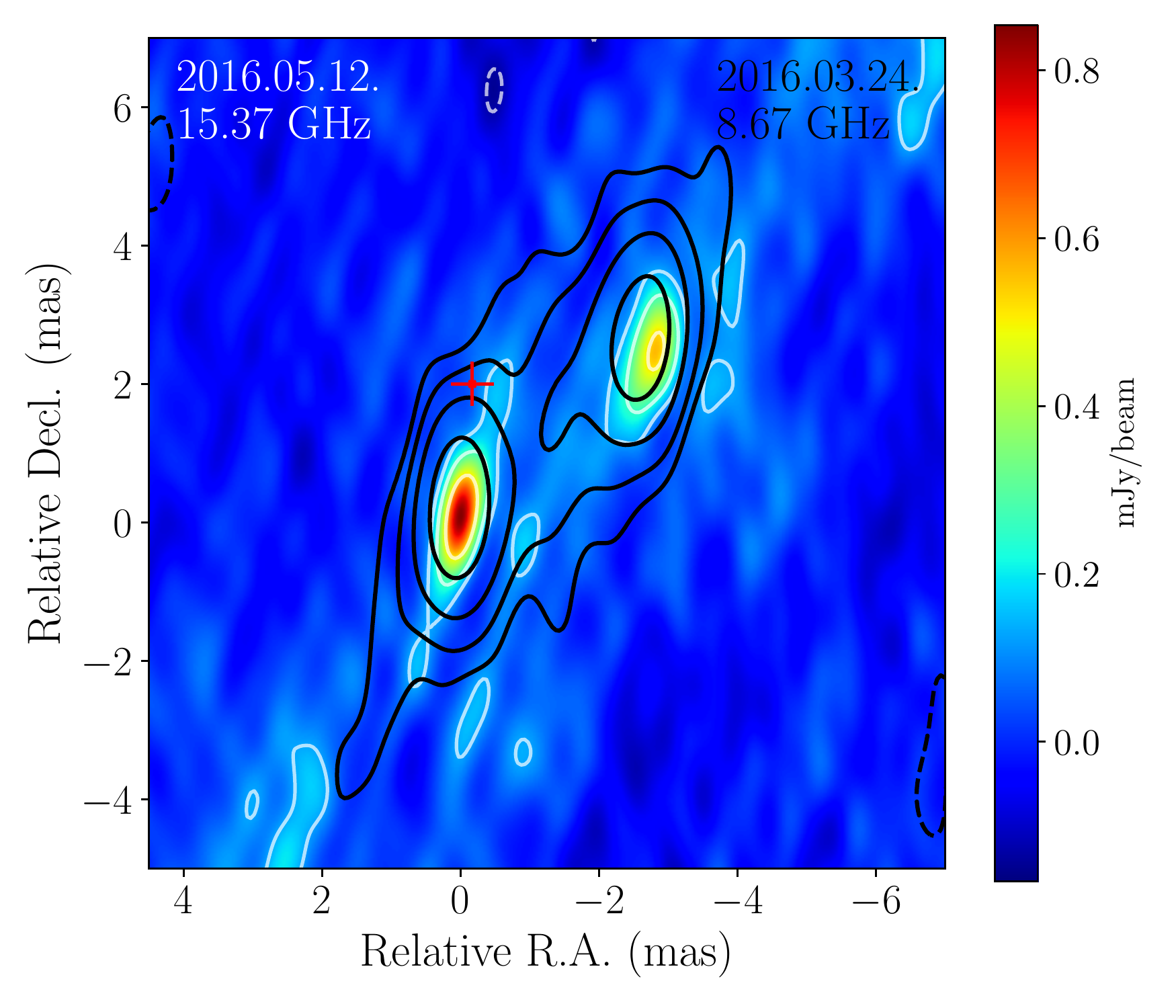}
    \caption{$15.37$-GHz VLBA image overplotted with the \textit{Gaia} position and its uncertainty (red cross). White contours are at $0.83$~mJy\,beam$^{-1}$ $\times$ ($-15$, $15$, $30$, $60$)\%, and the restoring beam size is $1.29\,\mathrm{mas} \times 0.48\,\mathrm{mas}$ at $\mathrm{PA}=-3.96\degree$. Black contours representing the $8.67$\,GHz source structure are at $1.56$~mJy\,beam$^{-1}$ $\times$ ($-7.5$, $7.5$, $15$, $30$, $60$)\%, and the FWHM of the restoring beam is $2.13\,\mathrm{mas} \times 0.86\,\mathrm{mas}$ at $\mathrm{PA}=-3.99\degree$.}
    \label{fig:gaia}
\end{figure}

Spectral index maps ($S_\nu\propto \nu^\alpha$) were created between our two VLA images at $1.5$ and $6.2$\,GHz, as well as between our $1.7$-GHz EVN image and the $4.38$-GHz VLBA image of \citet{mooley18}, plotted with black contours in Fig.~\ref{fig:gaia}. To align the images on the optically thin jet components, we used 2D cross-correlation.

\begin{figure}[h!]
    \centering
    \includegraphics[width=\linewidth]{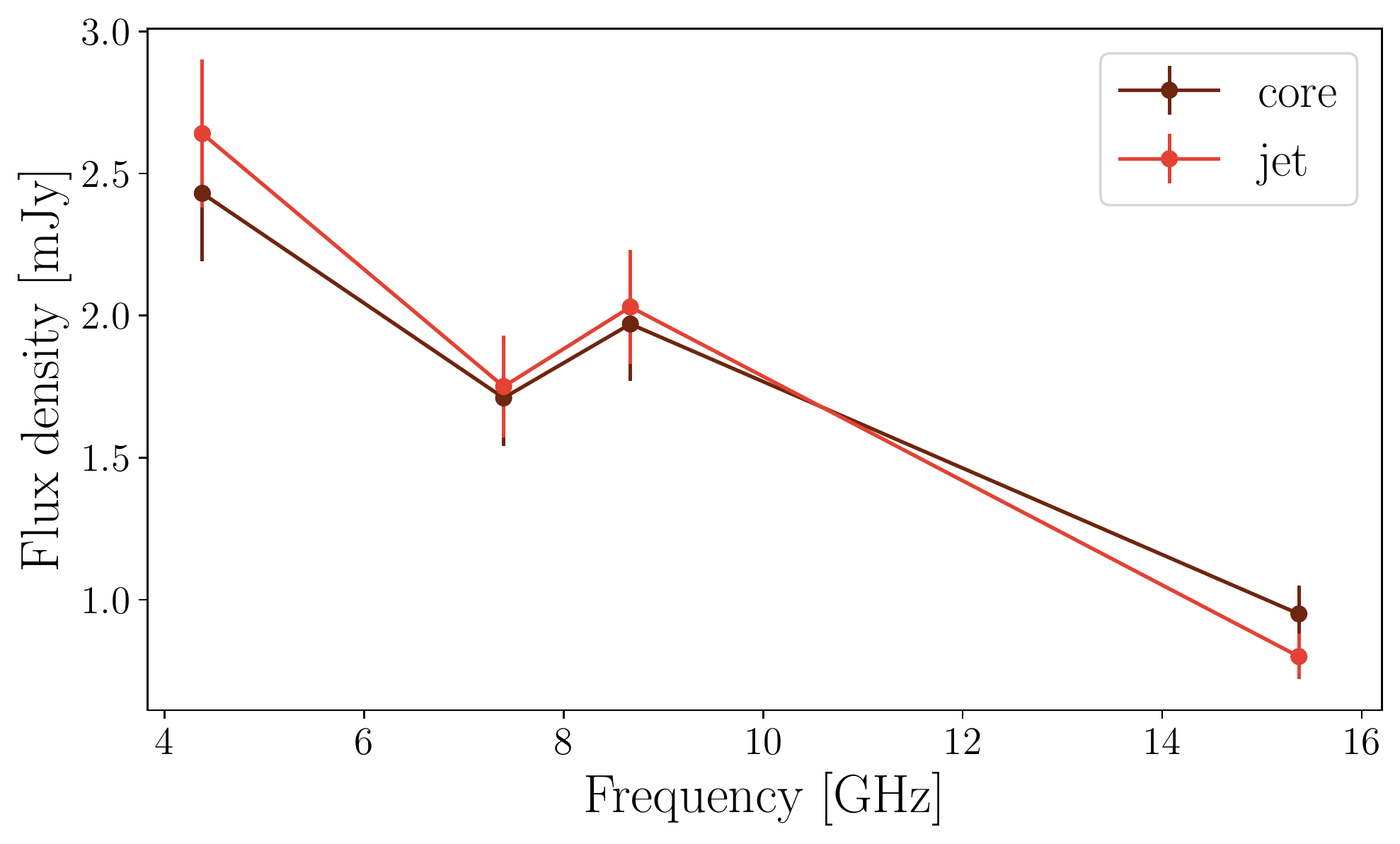}
    \caption{Spectra of the core (SE) and jet (NW) components of the VLBA images.}
    \label{fig:comps}
\end{figure}

\noindent The image pairs had the same restoring beam, map and pixel size, as well as the same minimum and maximum $(u,v)$ distance. The resulting spectral index maps are shown in Fig.~\ref{fig:spec}. Due to the low signal-to-noise ratio outside of the peaks of the core and lobe regions, we do not recover spectral index solutions there.

Spectra are flat in the VLA core, i.e. the central component, and the eastern lobe, however, $\alpha$ values indicating a steeper spectrum are measured in the western lobe (Fig.~\ref{fig:spec}). The flat spectrum of the eastern lobe might indicate a shock region, which is supported by the bending seen in the $6.2$-GHz VLA image (Fig.~\ref{fig:vla}), possibly happening due to the interaction with the ISM. The steeper spectrum of the western lobe, however, can be explained by an older population of electrons present in this region. The VLBI spectral index map created between $1.7$ and $4.38$~GHz shows a flat spectrum both for the core and the northwestern jet component. Since we assume that the amplitude calibration is accurate within $10\%$, spectral index errors are estimated to be $\pm0.15$. See also figure~$3$ of \citet{mooley18} for the spectral index map between $8.67$ and $15.37$\,GHz that show steep spectra in both VLBI components. We also plot the spectra of the core and jet components of the VLBI images in Fig.~\ref{fig:comps}, where both the core and the jet components show a steep spectrum between $1.7$ and  $4.38$\,GHz. 
In addition, the steepening towards the jet edge might be an artifact due to the low brightness of the component. While our results confirm the southeastern component as the core \citep{mooley18}, here we must note that the observations used to create the spectral index map were made half a year apart, so these results must be interpreted cautiously because of possible flux density variability.

The radio spectrum compiled in \citet{benke18}, as well as the broadband spectral energy distribution (SED) in \citet{foord17} show no deviation from single AGN spectra. The two models investigated by \citet{foord17} are the mini-disk and cavity models that represent different stages of the binary evolution and manifest in the SED as missing emission at different frequencies. However, they found that the optical to X-ray bands are well modeled with a composite non-blazar SED \citep{shang11} and the radio emission falls between what is expected from radio-loud and radio-quiet sources.

\begin{table*}[h]
      \caption[]{Characteristics of \texttt{modelfit} components.}
         \label{tab:comp}
         \centering
         \begin{tabular}{lccccc}
            \hline\hline
            \noalign{\smallskip}
            $\nu$ [GHz]\tablefootmark{a} & Array & Component & $S_{\mathrm{comp}}$ [mJy]\tablefootmark{b} & $b_{\mathrm{comp}}$ [mas]\tablefootmark{c} & $T_{\mathrm{b}}$ [$10^8$~K]\tablefootmark{d} \\
            \noalign{\smallskip}
            \hline
            \noalign{\smallskip}
            1.7 & EVN  &    & $7.15\pm0.72$ & $6.58\pm0.68$ & $2.2\pm0.5$\\
            4.38 & VLBA & SE & $2.43\pm0.24$ & $1.53\pm0.49$ & $2.0\pm1.3$ \\
                 &      & NW & $2.64\pm0.26$ & $1.23\pm0.49$ & $3.4\pm2.7$ \\
            7.40 & VLBA & SE & $1.71\pm0.17$ & $0.8\pm0.29$ & $1.8\pm1.4$  \\
                 &      & NW & $1.75\pm0.18$ & $0.9\pm0.29$ & $1.5\pm1.0$ \\
            8.67 & VLBA & SE & $1.97\pm0.2$ & $0.59\pm0.17$ & $2.8\pm1.6$ \\
                 &      & NW & $2.03\pm0.2$ & $0.83\pm0.17$ & $1.5\pm0.6$ \\
            15.37 & VLBA & SE & $0.95\pm0.1$ & $0.16$\tablefootmark{*} & $>6.0$\tablefootmark{*} \\
                  &      & NW & $0.8\pm0.08$ & $0.38\pm0.1$ & $0.9\pm0.4$ \\
            \noalign{\smallskip}
            \hline
         \end{tabular}
       \tablefoot{
       \tablefoottext{a}{Frequency of the observation}
       \tablefoottext{b}{Flux density of the component}
       \tablefoottext{c}{Size of the component (FWHM)}
       \tablefoottext{d}{Brightness temperature}
       \newline
       \tablefoottext{*}{Component is unresolved based on eq.~$2$ from \citet{kovalev05}}
       }
\end{table*}

\subsection{Brightness temperatures}

To study the brightness temperature of the compact radio emitting features in the VLBI images \citep[Figs.~\ref{fig:evn} and \ref{fig:gaia}, also][]{mooley18}, we fitted circular Gaussian components in the visibility domain with \texttt{modelfit} in \texttt{Difmap}. Characteristics of the \texttt{modelfit} components are summarized in Table \ref{tab:comp}. We assume flux density errors to be 10\% and the error in component size to be 20\% \citep{lister09}. We calculate the observed brightness temperature as
\begin{equation}
    T_{\mathrm{b, obs}} [\mathrm{K}] = 1.22\times 10^{12} \Bigg(\frac{S_{\nu}}{\mathrm{Jy}}\Bigg) \Bigg(\frac{\nu}{\mathrm{GHz}}\Bigg)^{-2} \Bigg(\frac{b_{\mathrm{min}}\times b_{\mathrm{maj}}}{\mathrm{mas}^2}\Bigg)^{-1} (1+z),
\end{equation}
where $b_{\mathrm{min}}$ and $b_{\mathrm{maj}}$ are the minor and major axis (FWHM) of the component. The resolution limit was computed based on eq.~$2$ in \citet{kovalev05}. For the one component for which the size fell below this limit, we can only calculate a lower limit of its brightness temperature. $T_{\mathrm{b, obs}}$ values fall between $10^7-10^9$~K. This confirms that the emission originates from AGN activity and not from spatially extended star formation in the host galaxy \citep{condon91}. The measured brightness temperatures are lower than the average core brightness temperatures of blazars in the MOJAVE 
 sample\footnote{\url{https://www.cv.nrao.edu/MOJAVE/}} of $1.39\times10^{11}$~K \citep{homan21}. This indicates that the emission of PSO J334 is likely not beamed and the jets are probably inclined at a large angle to our line of sight.

\subsection{VLBI and \textit{Gaia} astrometry}

Both the EVN and VLBA observations were carried out in phase-referencing mode, enabling precise relative astrometric measurements. We determined the position of the brightness peak in the VLBI images using \texttt{MAXFIT} in \texttt{AIPS}. Since we expect that the optical emission originates from the vicinity of the central engine, i.e., the accretion disk and the inner jet \citep{kovalev17}, we can use \textit{Gaia} data \citep{gaia} from the third data release \citep[DR3,][]{gaia-dr3} to identify the nucleus position. The standard error of the optical position is $0.27$~mas in right ascension and $0.32$~mas in declination based on \textit{Gaia} DR3. VLBI astrometric errors are comparable and estimated to be $0.3$~mas for low declinations with the VLBA based on the analysis of \citet{pradel06}. We also take into account the uncertainty of the position of the calibrator, J2217+0220. The source is in the 3rd realization of the International Celestial Reference Frame \citep[ICRF3,][]{charlot20}, and the formal position errors are $0.0765$~mas and $0.1137$~mas in right ascension and declination, respectively. We add these error estimates in quadrature, and estimate errors to be $0.31$~mas in right ascension and $0.32$~mas in declination. We find that the southeastern component at $8.67$~GHz is closer to the optical position of the source (see black contours in Fig.~\ref{fig:gaia}). We also see that the differences in right ascension are within the expected uncertainties, except at $4.38$\,GHz where $\sim 2$~mas offset was found between the \textit{Gaia} and the VLBI core positions. However, the optical--radio differences in declination reach up to $\sim 4$~mas, but show no frequency dependence.

We suspect that these differences in declination arise from uncorrected tropospheric and ionospheric effects. The contribution of the ionosphere to phase errors is significant at low frequencies, and since its rapid changes are difficult to model, we expect that this affects the measurement of the source position considerably. Based on $22$\,GHz VLBA observations, \citet{petrov23} determined that the ionospheric effects contribute around $0.1$~mas to the error budget both in right ascension and declination for northern hemisphere sources. However, for southern hemisphere targets, the accuracy of ionospheric corrections in declination is only $0.3$~mas. In addition, at low elevations, errors from the wet tropospheric component were shown to dominate over other systematic errors, and affect the precise measurement of source declination far more seriously than of the right ascension \citep{pradel06}. In order to be able to account for the ionospheric and tropospheric effects on VLBI astrometry more accurately, we need more extended studies such as in \citet{brisken02}, \citet{pradel06}, and \citet{petrov23}, to explore how these effects influence the error budget of our measurements for various VLBI arrays \citep[see e.g. the Long Basline Array and the VLBI Exploration of Radio Astrometry,][]{petrov19, honma08} and to develop methods to correct for them in the most effective way.

\section{Summary}
\label{sum}

\citet{liu15} originally proposed that PSO\,J334 might be a SMBHB based on the periodic variability observed in the optical light curves. The scenario was later disfavored as extended time baselines revealed no sinusoidal periodicity in the optical flux \citep{liu16}, and SED reconstruction did not show any signatures expected from a binary AGN system \citep{foord17}. The only remaining indication of a companion SMBH was the possible precession inferred from multi-frequency radio observations of \citet{mooley18}, where the large misalignment of $39\degr$ between the pc- and kpc-scale structures led the authors to reinstate PSO\,J334 as a SMBHB candidate.

Our high-resolution radio observations carried out with the VLA at $1.5$ and $6.2$\,GHz reveal a lobe-dominated quasar \citep[in agreement with the findings of][]{mooley18}, likely oriented at a large angle to our line of sight. Based on polarimetric imaging and the analysis of the arm length and brightness ratios, we suggest that the source is embedded in an asymmetric environment and similarly to the high-redshift radio galaxy 4C\,41.17 \citep{gurvits97}, its morphology is affected by the surrounding ISM, creating a bending in the eastern jet. PSO\,J334 also shows remarkably straight jets at 6.2\,GHz, that disfavor jet precession, the only argument left to support the SMBHB status of the object. In addition, our $1.7$\,GHz EVN observations confirm the significant misalignment between the outer lobes and the inner jet observed by \citet{mooley18}. We suggest that this misalignment can be explained by a newer phase of AGN activity where the orientation of the component ejection has changed. In conclusion, we find no evidence to support the binary nature of PSO\,J334, and it should be considered an ordinary quasar.

\begin{acknowledgements}
The authors would like to thank the anonymous referee for their useful suggestions. We thank J. Livingston for his valuable comments to the manuscript.
The EVN is a joint facility of independent European, African, Asian, and North American radio astronomy institutes. Scientific results from data presented in this publication are derived from the following EVN project code: RSF08.
The National Radio Astronomy Observatory is a facility of the National Science Foundation operated under cooperative agreement by Associated Universities, Inc.
This research was supported through a PhD grant from the International Max Planck Research School (IMPRS) for Astronomy and Astrophysics at the Universities of Bonn and Cologne. 
This work was supported by the Hungarian National Research, Development and Innovation Office (NKFIH, grant number OTKA K134213).

\end{acknowledgements}

%
%

\end{document}